# Andreev reflection in ferrimagnetic $CoFe_2O_4$/$SrRuO_3$ spin filters

Franco Rigato[1], Samanta Piano[2,3], Michael Foerster[1], Filippo Giubileo[3], Anna Maria Cucolo[3], and Josep Fontcuberta[1]

[1] *Institut de Ciència de Materials de Barcelona, CSIC, Campus UAB, Bellaterra 08193, Spain*

[2] *School of Physics and Astronomy, University of Nottingham, University Park, Nottingham, NG7 2RD, United Kingdom*

[3] *Physics Department and INFM-CNR SUPERMAT Laboratory, University of Salerno, Via S. Allende, 84081 Baronissi (SA), Italy*

We have performed point contact spectroscopy measurements on a sample constituted by a metallic ferromagnetic oxide ($SrRuO_3$) bottom electrode and a tunnel ferrimagnetic ($CoFe_2O_4$) barrier. Andreev reflection is observed across the tunnel barrier. From the comparison of Andreev reflection in $SrRuO_3$ and across the $CoFe_2O_4$ barrier we infer that the ferrimagnetic barrier has a spin filter efficiency not larger than +13%. The observation of a moderate and positive spin filtering is discussed in the context of the microstructure of the barriers and symmetry-related spin filtering effects.

## I. INTRODUCTION

The spin polarization is a key quantity for the spintronics. In the last few years, spin filters, constituted by a ferro(ferri)magnetic tunnel junction, have emerged as promising alternative to create artificial spin polarized current sources. Ferromagnetic spin filters rely on the spin-dependent transmittance of ferromagnetic tunnel barriers due to the existence of an exchange-split band gap. Following the pioneering work of Moodera et al.[1] who showed spin-filtering at low temperature using EuS ferromagnetic tunnel barriers, spin-filtering has been demonstrated using perovskite oxides[2] ($BiMnO_3$) and more recently, spinel oxides: $NiFe_2O_4$ (NFO)[3,4] and $CoFe_2O_4$ (CFO)[5] barriers. As the Curie temperature of ferrites is well above room temperature, efficient room-temperature spin-polarized sources could be obtained using these oxides. The most simple spin-filter structure is formed by two metallic electrodes: a non spin-polarized current source (M) and a ferromagnetic layer (FM), acting as a spin analyzer, with a ferro(ferri)magnetic tunnel barrier (FI) between them. Determination of the spin-filtering efficiency ($P_{FI}$) of the FI barrier has been achieved by measuring the magnetoresistance of the tunnel junction and using the Jullière model to derive $P_{FI}$. Using this approach low-temperature values of +22% and -25% have been reported for NFO[3,4] and CFO[5] respectively. However, by using the Meservey-Tedrow technique, positive $P_{CFO}$ values (ranging from +6% to 26% depending on film growth conditions) have been obtained in CFO-based spin filters[6]. Theoretical calculations of spin-dependent electronic structure of spinels indicate that the lowest energy conduction band is spin-down[7], thus predicting a $P_{FI} < 0$. It thus follows that, in spite of its relevance for further progress in spintronics, no definitive determination of $P_{FI}$ is yet available for spinel-based spin filters.

The Point Contact Andreev Reflection (PCAR) technique has been introduced as a tool to measure the spin polarization of carriers in ferromagnetic materials[8]. In the case of a normal metal/superconductor junction, an incoming electron from the normal metal with energy smaller than the superconducting gap cannot enter into the superconducting (SC) electrode and is reflected as a hole in the normal metal, simultaneously adding two electrons (a Cooper pair) to the condensate in the superconductor. This process, known as Andreev reflection (AR), causes an increase of the conductance around zero bias $G(V\approx0)$ compared to the conductance at voltages $G(V)$ well above the superconducting gap ($\Delta/e$) by a factor of two. Since the reflected hole is created in the density of states with opposite spin than the incoming electron, the AR process is partially suppressed when the respective densities of states are not equal, as in the case of a ferromagnetic metal/superconductor (FM/SC) interface. In particular, in fully spin-polarized metals, all carriers have the same spin orientation and the AR should be totally suppressed because a hole cannot be created in the opposite density of states. Thus, the absolute value (but not the sign) of transport spin polarization │P│ of a ferromagnetic material can be inferred from the grade by which AR is suppressed in a measured conductance spectra[8,9]. It has been recently predicted[10] that the insertion of a spin-filtering barrier to form a SC/FI/M structure should lead to the subsequent modification of the AR process by spin-selective tunnelling across the ferromagnetic insulator.

In this paper we report PCAR spectroscopy experiments realized by pressing a superconducting Nb tip on a spin filter constituted by a $CoFe_2O_4/SrRuO_3$ (CFO/SRO) thin film bilayer. We show that AR occurs across the ferrimagnetic CFO tunnel barrier of the $Nb/CoFe_2O_4/SrRuO_3$ (SC/FI/FM) structure demonstrating spin-preserved tunnelling through the CFO barrier. Analysis of the experimental conductance data by means of the modified BTK model[11] allows determining the polarization P of the CFO/SRO bilayer. PCAR experiments have also been conducted on the bare SRO layers, and we obtained P ≈ 42(1)%. These results indicate a very modest filtering efficiency of the CFO barriers (< +13%). Implications of these findings are discussed.

## II. SAMPLE PREPARATION AND PRE-CHARACTERIZATION

CFO/SRO bilayers have been deposited by RF magnetron sputtering from stoichiometric targets, on a single crystalline (111)$SrTiO_3$ substrate. The bottom electrode is a 25 nm thick ferromagnetic ($T_{Curie}$ ≈ 120 K) and metallic (ρ ≈ 200 μΩ•cm at 10K) SRO epitaxial film, deposited at ≈0.5 nm/min in a mixed atmosphere of argon and oxygen (ratio 3:2), with a total pressure of 100 mTorr; the substrate temperature was 725 ºC. The top layer is a 3 nm thick epitaxial CFO film, grown at ≈0.1nm/min with an $Ar/O_2$ ratio of 10:1, and a total pressure of 250 mTorr at 500 °C. The roughness of the SRO and the CFO layer are lower than 0.3 nm. The saturation magnetization of a bare 3 nm thick CFO film[12], measured by using a Quantum Design's SQUID, was ≈524 emu/cm$^3$. This value is somewhat larger than the bulk value. Similar enhancement had been reported earlier for ultrathin films of other spinel oxides[13] and attributed to partial cation inversion of the spinel structure and subsequent electronic and magnetic ordering modifications.

The bilayer structure was characterized by Conductive-Atomic Force Microscope (C-AFM), with a Nanotec Cervantes AFM, using Nanosensors tips (CDT-NCHR). Electrical measurements were performed in 2-point electrical configuration: the C-AFM probe was grounded, while the SRO bottom electrode was positively biased. C-AFM measurements were conducted in dry nitrogen atmosphere, using a feedback-force of ≈850 nN and an applied voltage of 800 mV.

In Fig. 1(a) we show a resistance map measured by C-AFM: at each point of the surface the system records the current between tip and sample under constant bias voltage. This image indicates a very high electrical homogeneity in a large area (3 μm x 3 μm). The simultaneously recorded topographic maps (not shown) confirmed the absence of particular defects and an extremely smooth surface (rms < 0.3 nm). The histogram of the resistance values, shown the Fig. 1(a)(inset) indicates a narrow distribution of log R (half-width at half-maximum ≈ 3.5%), centred around ≈8.25 (≈178 MΩ). As the tunnelling current depends exponentially on the barrier thickness, the data in Fig. 1(a) signals also an extremely small variation of the barrier thickness[14]. Subsequent measurements were repeated in the same area; neither the resistance map nor the corresponding topographic image evidenced significant changes in the surface properties other than a minor reduction of the overall resistance likely associated to residual surface contamination removal. No traces of indentation or scratching could be detected. Tests repeated at different locations on the sample surface yield very similar results thus confirming the homogeneity of the surface properties.

The Current-Voltage (*I-V*) curves measured at different points on the surface (one example is shown in Fig. 1b) show clear characteristics of tunnelling behaviour. The asymmetry visible in *I-V* curves is due to the asymmetric contact configuration for forward and reverse biasing of the tunnel structure and it is of no relevance for PCAR experiments, performed in the subgap region, to be described in the following.

### III. POINT CONTACT SPECTROSCOPY

Point contact junctions were formed by pushing a soft Niobium tip, obtained by mechanical cutting and chemical etching a Nb wire (diameter ≈ 0.2 mm), onto the CFO surface of the CFO/SRO bilayer. The experimental setup utilized for the measurements has been described elsewhere[15]. Due to the softness of Nb and to the hardness of the CFO surface, we do not expect significant scratching or penetration of the CFO film by the tip. The effective electric contact radius

*d* can be estimated by using the approximation[16] R = 4ρ*l* / 3π*d*² + ρ / 2*d* and employing the values of the resistance R measured at high bias (≈200-300 Ohm), the resistivity of equivalent SRO films (ρ ≈ 200 μΩcm) at 4 K and a mean free path *l* of about 15 Å, estimated from ρ[17]. It turns out that the estimated contact size *d* is of about 45-60 Å, implying that although the contacts are larger than the mean free path they are still in an intermediate regime *d* ≈ *l*.

We have recorded conductance curves at low temperature (T = 4.2 K) at different positions on the film and with different pressures between tip and sample. In Fig. 2 we show representative conductance spectra of two different types of junctions we measured: Junctions 1 and 2. Data have been normalized by using the background conductance estimated at large voltage (V >> Δ/e) regions, where Δ is the superconducting gap of Nb (Δ ≈ 1.5 meV). Both spectra display a clear increase of the conductance around zero bias suggesting that the transport is mainly due to the Andreev reflection process. Moreover, Andreev reflection, presumably spin-filter weighted, takes place across the ferrimagnetic barrier as predicted in Ref. [10]. From data in Fig. 2 it is clear that the normalized conductance G(≈0 V) < 2 thus indicating some suppression of AR as expected for a spin-polarized electron tunnelling.

It is noteworthy in Fig. 2 that the features in the conductance spectrum appear at energies that are sensibly higher than what is expected for Nb. This observation is commonly attributed to the presence of a spread resistance $R_S$ arising from the resistance of the sample between the junction and one of the measuring contacts[18]. The effect of $R_S$ is to shift the coherence peaks from V ≈ Δ to larger voltages and subsequently changing the G(0) values. Following Woods et al.[18] the spread resistance is included in our modelling of the PCAR curves by considering two contributions to the measured voltage[19]:

$$V(I) = V_{PC}(I) + V_S(I) \quad (1)$$

and so the measured conductance G(V) will be fitted by using

$$G(V) = \frac{dI}{dV} = \left(\frac{dV_{PC}}{dI} + \frac{dV_S}{dI}\right)^{-1} \quad (2)$$

$V_{PC}$ and $V_S$ are the voltage drops at the FM/FI/SC junction and at $R_S$, respectively. To calculate $V_{PC}(I)$ we used the modified BTK model[11] considering an effective P spin-polarized current coming out from the spin-filter. For simplicity, a ballistic regime (*l* >> *d*) of transport across the junction will be assumed. Its application to the present case, where *l* ≈ *d*, at first sight may seem problematic. However, as shown in Ref. [18], the potential errors introduced by applying ballistic formulas to diffusive contacts, as far as Z is not too small, have a negligible impact on the extracted P values and the uncertainty is translated into the barrier transparency Z. The other relevant fitting parameters are: the superconducting gap $\Delta$ and a smearing coefficient ($\Gamma$) that allows modelling of the broadening of the gap edges and inelastic scattering at the interfaces[20]. The local temperature has been assumed to be the same like measured at the sample holder, which is immersed in a liquid He bath.

The spread resistance $R_S$ was determined beforehand by setting a gap value $\Delta$ equal to the BCS value for Nb, thus eliminating a possible degeneracy in $\Delta$ and $R_S$. Consequently, $\Delta$ cannot be considered a completely free parameter, but it is set close to the BCS value. In the subsequent fitting step, $R_S$ was kept fixed to the determined value and P, Z, $\Delta$ and $\Gamma$ were allowed to vary. The solid line in Fig. 2a is the result of the optimal fit of $G(V)$ using P ≈ 39%, Z ≈ 0.13, $\Gamma$ ≈ 0.22 meV, $\Delta$ ≈ 1.50 meV and $R_S/R_{PC}$ ≈ 0.5. The low interface barrier transparency Z used to fit our experimental data indicates that our measurements are not significantly affected by a possible dependence of the values of P on Z[21-23]. The quality of the fit and its robustness on variations of P can be better appreciated in Fig. 2c where we show the low-voltage zoom of data in Fig. 2a. In Fig. 2c we also include fits obtained by fixing $\Delta$ (1.5 meV) and P to be larger (45%) or smaller (35%) than P from optimal fit of Fig. 2a and allowing Z, $\Gamma$ to vary. We notice in Fig. 2c that fixing larger (smaller) P values lead to smaller (larger) $\Gamma$ values as predicted in Ref. [22], just illustrating that moderate inelastic interface scattering ($\Gamma$) decreases the AR probability. Thus the $\Gamma$ and P parameters in the AR spectra mix together and distinction among both effects is challenging[20]. In spite of this, it is clear from these data that $P_{J1}$ should be in the 45-35 % range for Junction 1.

To treat a possible degeneracy of fits to PCAR spectra, it has been proposed[25] to perform fits of ($\Delta$, Z, $\Gamma$) for different fixed $P_{trial}$, and check the resulting sum of the squared deviations $\chi^2$ of the fits as function of $P_{trial}$. In Fig. 3, results of such analysis are shown for the data of Junction 1. A clearly defined minimum of $\chi^2$ at about 39(1)% is found, which falls well within the previously determined range of values, thus we conclude that $P_{J1}$ ≈ 39(1)%. However, for the data of Junction 2 (Fig. 2b, which will be discussed below), such procedure proved to be ill-defined when using the proximity reduced gap, due to the larger number of fit parameters.

It has been demonstrated[25] that an error in the normalization of the conductance spectra can lead to false minima in $\chi^2$ vs. $P_{trial}$. Therefore the $\chi^2$ analysis was repeated using a deliberately wrong normalization of the data in Fig 2a (divided by an additional factor n = 1.001). Still, $\chi^2$ reproduces the minimum but values of $\chi^2$ are considerably enlarged. For normalization just slightly smaller than 1 (factor 0.999) reasonable fits to the data are no longer possible and values of $\chi^2$ increase drastically, demonstrating the validity of the original normalization.

In Fig. 2b we show an example of a $G$(V) curve measured in some different contacts (Junction 2). Data in Fig. 2b display characteristic features of AR reflection but also the occurrence of proximity effects and subsequent smaller gap formation ($\Delta_P$)[11,24]. As CFO is a ferromagnetic insulator it may be supposed that a gap reduction may occur in some part of the tip, likely due to the stray field of the ferrimagnetic barrier. The modified BTK model has been worked out including explicitly two gaps ($\Delta$ and $\Delta_P$)[11]. We used the corresponding expressions to fit the data in Fig. 2b. The solid line is the result of the optimal fit of $G$(V) using, $\Delta \approx 1.48$ meV, $\Delta_P \approx 0.99$ meV, $\Gamma \approx 0.0$ meV and $R_S/R_{PC} \approx 0.36$. The values Z ($\approx 0.20$) and P (31%) determined for this contact are quite similar to those extracted from data of Junction 1 (Fig. 2a). In the zoom of the low-voltage region (Fig. 2d) we include the fits obtained by fixing P to larger (35%) or smaller (25%) values while keeping $\Delta$ and $\Delta_P$ constant. The small $\Gamma$ values obtained from the best fits ($\Gamma \approx 0.1$), most probably result from the fact that the broadening of the spectrum is captured during the fit by the presence of the two gaps required to fit the dips. Although the available data do not allow to disentangle both contributions, it is clear from these data that P $\approx$ 31(3)% is a robust result for Junction 2. At this point it is worth to recall that dips in PCAR conductance spectra, as they appear in the data of Junction 2 have also been explained as arising from the superconductor reaching its critical current and thus adding some normal conducting state finite resistance[26]. Although the origin of the dips in the conductance spectra cannot be decided with certainty, a smaller superconducting energy gap is in principle, consistent with the observation of critical current effects in Junction 2.

Therefore, it follows that Andreev reflection has been observed across a ferrimagnetic tunnel barrier (Fig. 4). From the analysis of the PCAR we infer that representative values of the effective polarization of our CFO/SRO spin filter (3 nm thick) are of about P $\approx$ 31(3)% and 39(1)% as determined for Junctions 1 and 2 respectively. To progress further and to extract the spin-filter efficiency of the CFO barrier requires the knowledge of the spin-polarization of electrons emitted from the SrRuO$_3$ electrode ($P_{SRO}$).

Significantly different $P_{SRO}$ results have been reported: Worledge and Geballe[27] reported Meservey-Tedrow type measurements of tunnel junctions having SRO as electrode, and determined $P_{SRO}$ = -9%; Nadgorny et al.[23], using PCAR with Sn tips, inferred a much larger value ($|P_{SRO}| \approx 53\%$). Although this strong discrepancy is not fully understood[23,27], consensus exist that

$P_{SRO} < 0$. Negative spin-polarization arises from the difference of Fermi velocities for spin-up and spin-down electrons emerging from SRO rather than the density of states at the Fermi level (which is practically identical)[27].

To obtain a more reliable basis for assessing the PCAR results from the CFO/SRO structure, additional experiments were performed on a SRO film equivalent to the bottom electrodes used in the bilayers. A representative PCAR spectrum of the SRO film is shown in Fig. 5. The polarization of SRO was again determined by fitting (solid line) to the modified BTK model and resulted in $P_{SRO} = 42\%$, somewhat lower but still in agreement with previously reported PCAR results within their variations[23]. In the inset of Fig. 5, the resulting $\chi^2$ for fits with various $P_{trial}$ to the SRO data with a clear minimum at $P_{SRO} = 41.5(1)\%$ is shown.

## IV. DISCUSSION

We are now in position to compare the effective polarization measured in the spin filter and the bare SRO electrode. One first notes that the values of P extracted from both junctions are smaller than that of the bare SRO electrode. Due to the fact that $P_{SRO}$ is recognized to be negative ($P_{SRO} < 0$), this observation implies that the spin polarization of the CFO barrier must be positive ($P_{CFO} > 0$). Notice that although PCAR experiments of a single ferromagnetic layer do not allow to extract the sign of the spin polarization, this is possible in the present structure, where two ferromagnetic layers are involved, if the sign of the spin polarizations of one of the layers (SRO in this case) is known.

It has been theoretically predicted that the lowest energy barrier in the exchange-split gap of CFO is the spin-down[7]; in such circumstances it could be expected that the spin-down channel is dominating ($P_{FI} < 0$) the AR (Fig. 4), and thus one could anticipate an effective P in CFO/SRO bilayers larger than in the PCAR of the bare SRO electrode. Experimentally this is not the case as $P_{J1}$ and $P_{J2}$ are found to be smaller than $P_{SRO}$. We will discuss below on this discrepancy.

We next consider the values of the spin filtering efficiency of the CFO layers. The two extracted values of $P_{J1, J2}$ ($\approx 39(1)\%$ and $\approx 31(3)\%$) represent two distinct situations. Whereas the second value (Junction 2) would indicate a substantial spin filtering of CFO, this is not so for data from Junction 1, where a rather small different with the bare SRO electrode is observed.

We can define the spin-filter efficiency of the CFO barrier $P_{CFO}$ by[28]:

$$P = \left| \frac{P_{SRO} + P_{CFO}}{1 + P_{SRO} P_{CFO}} \right| \qquad (3)$$

where P is the effective spin polarization measured in the PCAR experiment. Using $P_{SRO} = -42\%$, the two solutions of Eq. (3) are $P_{CFO} \approx +4\%$ and $+73\%$ for $P_{J1}$ and $P_{CFO} \approx +13\%$ and $+67\%$ for $P_{J2}$. Among these two sets of possible values, which physically arise due to the fact that PCAR can not determine the sign of the measured spin polarization, the largest pair of values (73% and 67%) are most likely inappropriate as they could lead to spin filtering efficiencies for CFO much larger than reported values[5,6]. On the other hand, we notice that $P_{CFO} \approx +13\%$, as determined for Junction 2, is within the range of Meservey-Tedrow results ($+6\% \leq P_{CFO} \leq +26\%$, depending on the growth conditions)[6] obtained on CFO barriers of nominally equal thickness ($d \approx 3$ nm). Notably, the sign of the effect agrees with the Meservey-Tedrow experiments, which is, however, different from the one determined in tunnel magneto resistance measurements using $Al_2O_3$ tunnel barriers. On the other hand data for Junction 1 appears to indicate a very marginal spin filtering effect. This could be related to the fact that the CFO barrier is locally suppressed either by a mechanical effect associated to the Nb-tip pressure (although, as mentioned above, scratching effects have not been observed when using the harder C-AFM tip) or a locally poorer homogeneity of the insulating barrier. Although the CFO layers appears homogenous and robust in CAFM measurements, the possibility that the Nb tip penetrates the CFO layer either through pinholes or completely when crushed onto the sample cannot be excluded. Under such circumstances, it turns out that $P_{J2}$ and thus $P_{CFO} \approx +13\%$ constitutes the most representative value of the spin filtering efficiency of the present CFO barriers. It could be argued that a similar result would be obtained if the CFO barrier does not spin-filter at all but only contributes to depolarize the electron current from SRO. However, the experimental observation of spin-filtering in CFO in tunnel structures[5,6] does not support this view.

Before concluding we would like to comment on the positive sign observed for $P_{CFO}$. As mentioned, this observation is opposite to theoretical predictions[7], which are based on the electronic configuration of an ideally inverse spinel structure of CFO where all $Co^{2+}$ occupy the octahedral sites of the unit cell and the $Fe^{3+}$ equally populate the octahedral and tetrahedral sites. However, magnetization data of nanometric thin films[13,29] of various spinels, including $CoFe_2O_4$ and $NiFe_2O_4$ have provided conclusive evidence that the cationic distribution in nanometric thin films may largely differ from their bulk counterparts. As shown by the calculations of Szotek et al.[7], the insulating gap of CFO closes by some 75% when the cationic distribution is not that of the ideal inverse spinel structure but a normal one. It thus follows that the tunnel transport may be overcome by other non-spin preserving transport channels and therefore a reduced spin filtering efficiency could be anticipated for cationically-disordered films. To what extent the change of sign of the spin filtering efficiency is related to the same effect or not is not definitely settled. We also notice that the spin filtering efficiency should not be simply related to the exchange splitting of the insulator but the symmetry of the relevant wave functions may also play a role.

## IV. SUMMARY

In summary we have shown evidence that Andreev reflection occurs at ferro(ferri)magnetic tunnel barriers. Data collected using point contact spectroscopy allowed to estimate the effective spin-polarization of a current through the interface of a spin-filter and a superconducting tip. It turned out that a possible spin filtering effect of the spinel oxide $CoFe_2O_4$ tunnel barrier is limited to about +13% with the accuracy of these measurements. Observation of a positive spin filtering efficiency is unexpected and may suggest the relevance of spin-dependent orbital symmetry effects on the tunnel probability in spin filters. To the best of our knowledge this issue has not been theoretically addressed yet.

## ACKNOWLEDGEMENTS

Financial support from the Ministerio de Ciencia e Innovación of the Spanish Government Projects (MAT2008-06761-C03 and NANOSELECT CSD2007-00041) and from the European Union [ProjectMaCoMuFi (FP6-03321) and FEDER and Marie Curie IEF Project SemiSpinNano] is acknowledged. S.P. thanks B.L. Gallagher and C. J. Mellor for hosting the final stage of this research.

**Figure Captions**

Fig. 1: (Color online) C-AFM analysis of a CFO (3 nm)/ SRO (25 nm) bilayer on (111)STO: (a) resistance map in logarithmic color scale; inset: histogram of the resistance values over the surface. (b) I-V characteristic, with a schematic of the measurement circuit (inset).

Fig. 2: (Color online) Measured conductance spectra: (a) Junction 1; (b) Junction 2; the continuous line represents the best fits. (c), (d) Zoom around low bias; for Junction 1 and 2 respectively: Comparison of the best fits with simulations achieved forcing P to slightly different values, demonstrating the accuracy of the obtained polarizations.

Fig. 3 Characteristics of fits obtained for data from Junction 1. $\chi^2$ as function of $P_{trial}$ for standard normalization (n = 1) and varied normalization (n = 1.001).

Fig. 4: (Color online) Schematic of the density of states *vs.* energy of a SC-FI-FM (Nb-CFO-SRO) structure. The Andreev reflection (AR) process is illustrated (solid arrow). An incoming electron from the SRO side is reflected as a hole in the spin-reversed density of states, while a Cooper pair is added to the superconducting condensate. Parallel to AR, electron tunnelling into thermally excited states may occur (dashed arrow). For simplicity, only spin-down electrons are depicted and low-lying spin-down barrier has been assumed.

Fig. 5 (Color online) Measured conductance spectra of a SRO thin film, equivalent to the base electrodes used in CFO/SRO bilayers; the continuous line represents the best fit. Inset: $\chi^2$ as function of $P_{trial}$ for fits to the data in the main panel.

**Figure 1**

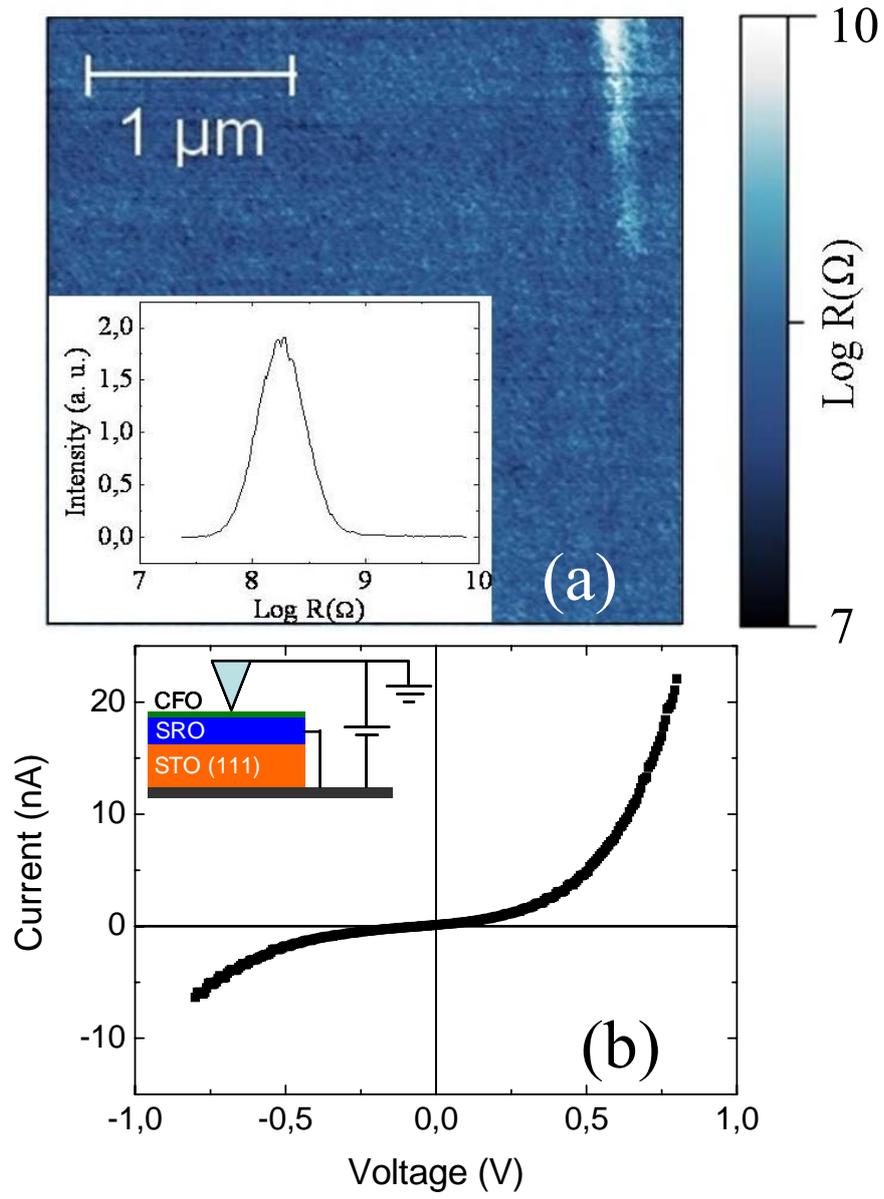

**Figure 2**

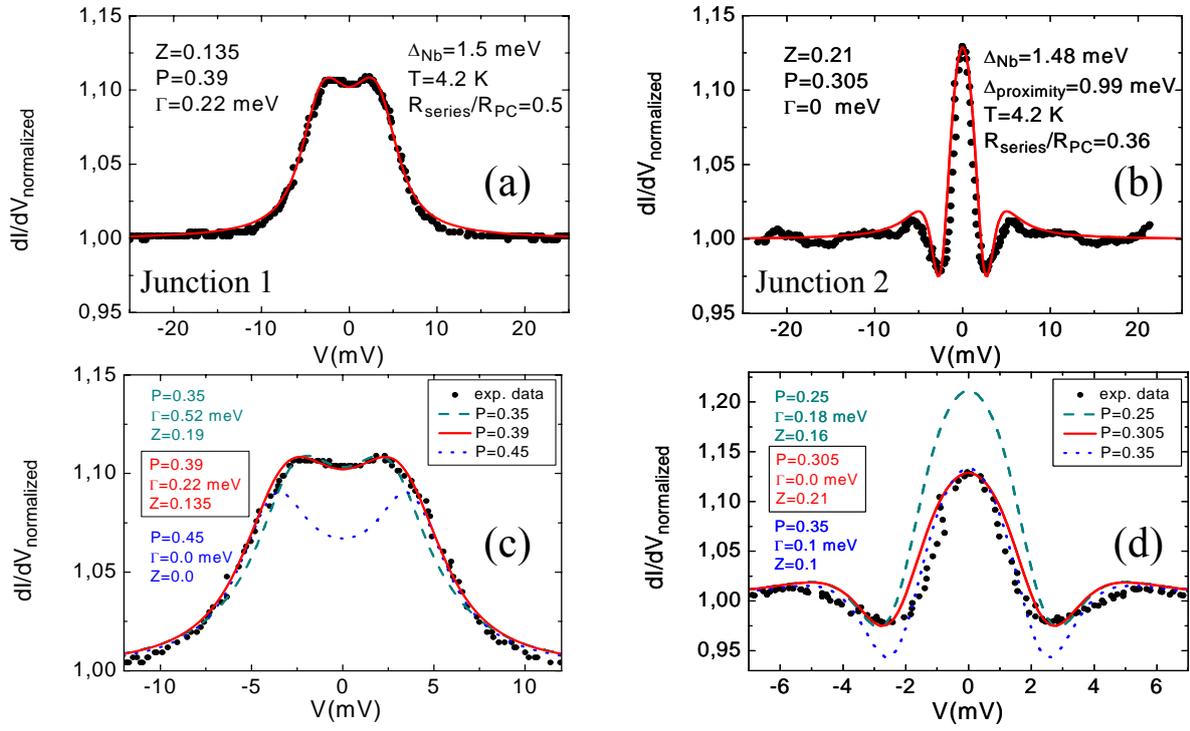

**Figure 3**

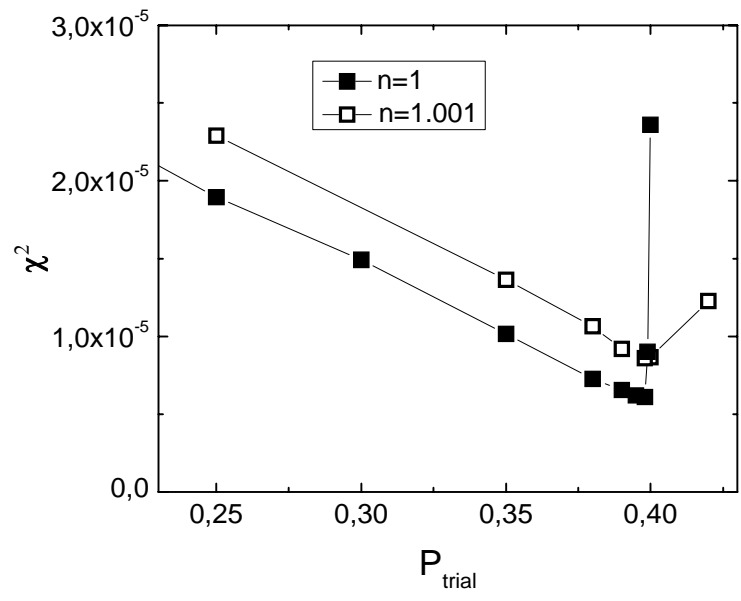

**Figure 4**

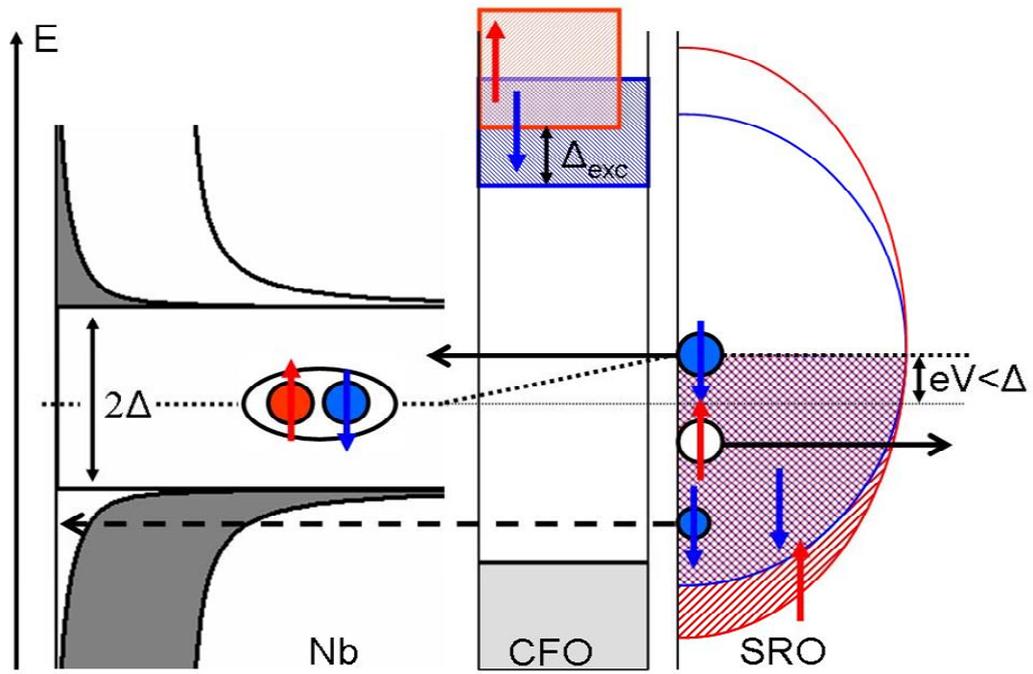

**Figure 5**

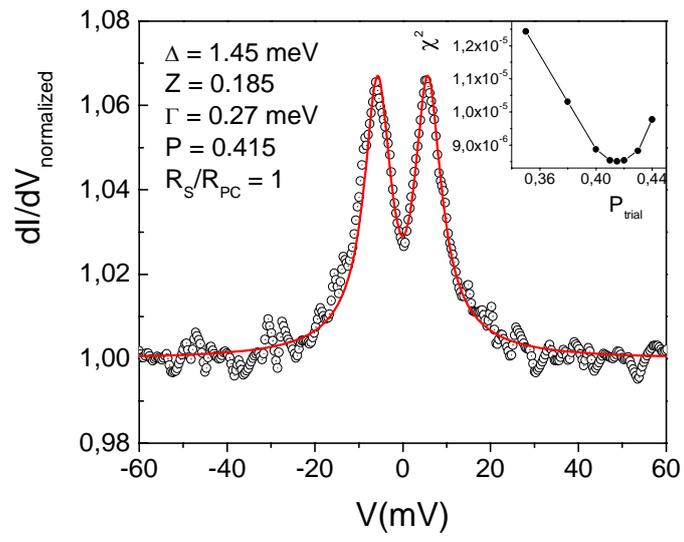